\documentclass[aps,prd,showpacs,twocolumn,floatfix,nofootinbib]{revtex4}
\usepackage{bm}
\usepackage{amssymb}
\usepackage{graphicx}
\usepackage{epstopdf}

\usepackage{graphicx}
\usepackage{dcolumn}

\begin{document}

\title{Self--forced gravitational waveforms for Extreme and Intermediate mass ratio inspirals}

\author{Kristen A.~Lackeos$^1$ and Lior M.~Burko$^{1,2,3}$}
\affiliation{$^1$ Department of Physics, University of Alabama in Huntsville, Huntsville, Alabama 35899, USA
\\
$^2$ Universit\'e d'Orl\'eans, Observatoire des Sciences de l'Univers en Region Centre, LPC2E Campus CNRS, 45071 Orl\'eans, France\\
$^3$ Department of Physics, Chemistry, and Mathematics, Alabama A\&M University, Normal, Alabama 35762, USA}
\date{June 8, 2012; Revised \today}
\begin{abstract}
We present the first orbit--integrated self force effects on the gravitational waveform for an I(E)MRI source. We consider the quasi--circular motion of a particle in the spacetime of a Schwarzschild black hole and study the dependence of the dephasing of the corresponding gravitational waveforms due to ignoring the conservative piece of the self force. We calculate the cumulative dephasing of the waveforms and their overlap integral, and discuss the importance of the conservative piece of the self force in detection and parameter estimation. For long templates the inclusion of the conservative piece is crucial for gravitational--wave astronomy, yet may be ignored for short templates with little effect on detection rate. We then discuss the effect of the mass ratio and the start point of the motion on the dephasing. 
\end{abstract}

\pacs{04.25.-g, 04.25.dg.Nk, 04.25.Nx, 04.70.Bw, 97.60.Lf}
\maketitle

\section{Introduction and Summary}

The detection of gravitational waves (henceforth GW) and the onset of the new field of gravitational-wave astronomy is one of the most exciting challenges for science in the XXI century, completing what is sometimes alluded to as Einstein's Unfinished Symphony. The detection of GW will open a new window onto the universe, that in addition to revealing exciting information on exotic systems such as black holes or cosmic strings is expected to also unravel as yet unexpected sources. 

One of the interesting sources for low--frequency GW are the so called I(E)MRI sources, or Intermediate (Extreme) Mass Ratio Inspirals. Those are the GW emitted by a system including a smaller compact object whose orbit decays into a much larger massive black hole (MBH). Typical sources are stellar mass black holes inspiraling into a supermassive black hole, like those residing at the center of galaxies, and also IMBHs (intermediate mass black holes) inspiraling into MBHs. The importance of such sources is that because of the extreme mass ratio  the smaller compact object can be viewed as a test particle, thus probing the spacetime of the larger black hole and its surroundings. {\it Inter alia}, such sources will allow us to test directly the Kerr hypothesis, and allow us to map the spacetime surrounding such exotic objects. Moreover, the detection of I(E)MRIs will allow us to determine the mechanisms that shape stellar dynamics in galactic nuclei with unprecedented precision \cite{Amaro-Seoane2012}. 

The orbits of I(E)MRIs are typically highly relativistic, and exhibit exciting phenomena, e.g.~extreme periastron and orbital plane precessions. Because the orbital evolution time scale (``radiation reaction time scale'') is much longer than the orbital period(s), over short time scales the orbit is approximately geodesic, yet on long time scales it deviates strongly from geodesic motion of the background. Instead, the smaller objects moves along a geodesic of a perturbed spacetime. Alternatively, one may construe the orbit as an accelerated, non-geodesic motion in the spacetime of the unperturbed central object, where the acceleration is caused by the self force (henceforth SF) of the smaller object \cite{Poisson-Pound-Vega}. 

Detection and parameter estimation of GW from  I(E)MRIs relies on the construction of theoretical templates. A number of approximation schemes for such templates are available. Firstly, the energy balance approach (``the radiative approximation'') uses balance arguments for otherwise conserved quantities, and relates the flux in these quantities to infinity and down the event horizon of the black hole with the particle's orbit, so that the latter can be adjusted to agree with the fluxes \cite{Hughes}. 
As the orbital evolution time scale is typically much longer than the orbital period(s), the radiative approximation is very satisfactory during the adiabatic phase of the motion. As the particle's orbit is affected by the fluxes away from it, when the orbit is not stationary one encounters complex retardation effects. 
Most currently--available EMRI waveforms have been obtained by such an approach. This approach, however, ignores conservative effects that do not register in the constants of motion.  

These retardation effects are completely avoided when one considers a local approach to orbital evolution in terms of the SF. (One should bear in mind, however, that the SF itself is a non-local quantity, with contributions arising from the quasilocal neighborhood of the particle and possibly beyond \cite{Anderson}.) In addition, the local approach to the calculation of orbital evolution via the SF is not restricted to the adiabatic regime, it avoids the complications associated with the rate of change of the Carter constant, and, most importantly, it includes also conservative effects that are discarded when one uses balance arguments. Over the last decade much progress has been made in the computation and understanding of the SF in General Relativity (for recent reviews of the self force in General Relativity see \cite{Poisson-Pound-Vega,barack:2009}).

The computation of the fully relativistic SF allows one to include conservative effects in the waveform templates, and study the importance of the conservative effects. True self consistent orbit and waveforms include the instantaneous solution of the coupled SF integrated equations of motion and the perturbation equations, or equivalently the interaction of the particle with its own field over its half--infinite past world line \cite{gralla-wald}. Very recently, for the scalar field toy model, such self consistent Schwarzschild orbits and waveforms were presented \cite{diener}. 
Here, we are making the simplifying assumption that the effects of the difference between the SF that is calculated for the actual orbit (the self consistent approach \cite{Pound,gralla-wald}) and that which is calculated for a geodesic of the same instantaneous orbital parameters, is smaller than the effects of the latter and hence negligible at first order. This approximation is valid for as long as the orbital evolution is adiabatic, that is as long as the orbital evolution time scale is much longer than the orbital period(s). In a Schwarzschild background of mass $M$, the adiabatic approximation holds when the mass ratio $\eta:=\mu/M$ is such that $\varepsilon\gg \eta^{1/2}$, where $\varepsilon$ measures the distance to the innermost stable orbit, specifically $\varepsilon=p-6-2\epsilon$ where $p$ is the semilatus rectum and $\epsilon$ is the orbital eccentricity \cite{cutler}.  In practice, our approximation is to a leading order in $\eta$ beyond geodesic motion. We neglect terms that are linear in second--order SFs, although our method is amenable to their inclusion when they become available. This approximation is valid for at least a part of the relevant parameter space \cite{burko04}, but as their inclusion would contribute linearly to the dephasing, the contribution of the conservative piece of the SF (hereafter CSF) may be isolated as is done here.  
Using true self consistent waveforms will both produce more accurate waveforms, and allow us to test the accuracy of this approximation. Most importantly, our approach allows us to see for the first time the effect of the CSF on GW emitted from IMRI sources. 

We present here the first waveforms obtained with inclusion of the CSF, and study its effect within the simple class of quasi--circular orbits around a Schwarzschild black hole.  Specifically, we study the effect of the system's mass ratio on the dephasing that occurs when one neglects the CSF. We find weak dependence of the dephasing on the mass ratio, in accord with expectations based on the scaling of the number of orbits with the inverse of the mass ratio, and the scaling of the dephasing effect of the CSF per orbit with the mass ratio. We also find that the dephasing depends quadratically on the initial point of the motion for the range of parameters we tested. 
We reiterate that second--order dissipative effects are ignored in this Paper. Their inclusion will guarantee the full consistency of the model, and will be comparable to the self-consistent approach. The inclusion of the second-order dissipative effects awaits further development to both theory and computational techniques. 

The organization of this Paper is as follows: In Section \ref{method} we discuss the computational and numerical methods that we use. In Section \ref{results} we discuss our results for the orbits (\ref{orbit}), the waveforms (\ref{waveform}) and the dependence of the dephasing on the mass ratio and the initial point of the motion (\ref{varying}). 

\section{Method}\label{method}

We use the fully relativistic SF obtained by Barack and Sago \cite{Barack-Sago} for circular Schwarzschild geodesics to drive the orbital evolution (our computation allows for an easy replacement with a different force expression, say one that includes second order dissipative effects, or spin--orbit coupling effects when becoming available), and compare the resulting waveforms with those obtained from the energy balance approach and those obtained when only dissipation is left in the SF, setting by hand the CSF to vanish. In practice, we consider a point source $\mu$ in a quasi--circular orbit around a Schwarzschild black hole $M$ with a mass ratio $\eta:=\mu /M$, and the motion starts at  a value of the semilatus rectum $p_0$ until it gets close to the ISCO at $p=6$. Such sources are relevant to the NGO capabilities: NGO will have the capability to detect GW emitted by an IMBH in the mass range $10^{2-4}M_{\odot}$ spiraling into a MBH in the mass range $3\times 10^5-10^7M_{\odot}$ such that the mass ratio is $10^{-3}-10^{-2}$ out to cosmological redshift $z\sim 2-4$. In addition, advanced LIGO could detect compact stellar sources spiraling into an intermediate mass black hole (IMBH) in the same mass ratio range \cite{brown:2007}. We specialize below to this mass ratio range, $\eta\in [10^{-3}-10^{-2}]$. Although the linearized  approach used here is intended to be used only when $\eta\ll 1$ and one may not simply extend the range of $\eta$ to high values and still expect accurate results, the error involved from neglecting $O(\eta^2)$ terms in the self force (specifically its dissipative piece) is comparable to the accuracy of our computation. In this sense, to within the accuracy of our numerics, we are justified in studying the IMRI case as long as we do not raise $\eta$ beyond $10^{-2}$. 

We integrate the equations of motion using the Barack--Sago SF which was calculated for momentary circular geodesics. As the Barack--Sago SF is tabulated for a select choice of orbital radii, we interpolate to intermediate values such that the original accuracy is maintained. Specifically, we match two asymptotic expansions: at large distances (which we take in practice to be $r>8M$) we take the standard post--Newtonian expansion for the luminosity in gravitational waves, and construct from it the temporal component of the SF. To 5.5 PN order the PN expression does not provide us with sufficient accuracy to reproduce all the $r\ge 8M$ data points of the Barack--Sago data. We therefore add an effective remainder term that appears like a 6PN term, and fit its two free parameters to agree with all the large distance tabulated data to all significant figures. The radial component, or CSF, is modeled by a PN--like expansion with four free parameters. At short distances we expand the SF such that convergence is fast and only four free parameters are needed for either component. Specifically, we expand the radial component about the ISCO, the Innermost Stable Circular Orbit at $p=6$. The expansion functions are as follows:

 \begin{eqnarray}
f^t_{r\le 8\,M}&=&-
\frac{1}{\sqrt{1-\frac{3M}{r}}\left(1-\frac{2M}{r}\right)} \; \bigg(\frac{M}{r}\bigg)^5\; \bigg(\frac{\mu}{M}\bigg)^2\;
\Bigg[a^-_0\nonumber\\
&+&a^-_1\,\frac{M}{r}+a^-_2\,\bigg(\frac{M}{r}\bigg)^2+a^-_3\,\bigg(\frac{M}{r}\bigg)^3 + \cdots \Bigg]
\end{eqnarray}
\begin{eqnarray}\label{pn}
f^t_{r\ge 8\,M}&=&-
\frac{32}{5}\,\frac{1}{\sqrt{1-\frac{3M}{r}}\left(1-\frac{2M}{r}\right)} \;\left(\frac{M}{r}\right)^5\;\left(\frac{\mu}{M}\right)^2\\
&\times&\;
\left[ {\rm PN}_{5.5}+\left(a^+_6+a^+_{6L}\,\ln\frac{M}{r}\right)\left(\frac{M}{r}\right)^6 + \cdots \right]\nonumber
\end{eqnarray}
\begin{eqnarray}
f^r_{r\le 8\,M}&=&
\bigg(1-\frac{2M}{r}\bigg)\; \bigg(\frac{M}{r}\bigg)^2
\bigg(\frac{\mu}{M}\bigg)^2\nonumber\\
&\times&\;
\Bigg[b^-_0+b^-_1\,\bigg(1-\frac{6M}{r}\bigg)+b^-_2\,\bigg(1-\frac{6M}{r}\bigg)^2\nonumber\\
&+&b^-_3\,\bigg(1-\frac{6M}{r}\bigg)^3 + \cdots\Bigg]
\end{eqnarray}
\begin{eqnarray}
f^r_{r\ge 8\,M}&=& \bigg(\frac{M}{r}\bigg)^2
\bigg(\frac{\mu}{M}\bigg)^2\;
\Bigg[b^+_0\\
&+&b^+_1\,\frac{M}{r}+b^+_2\,\bigg(\frac{M}{r}\bigg)^2
+b^+_3\,\bigg(\frac{M}{r}\bigg)^3 + \cdots\Bigg]\nonumber
\end{eqnarray}
where ${\rm PN}_{5.5}$ stands for the standard $\frac{11}{2}$--post--Newtonian expression (converting Eq.~(3.1) in \cite{tanaka:1996} from luminosity to $f^t$ -- see Appendix \ref{PN}). Fitting the free parameters, we find the values appearing in Table 1.
\begin{table}[htdp]
\caption{The fit parameters for the self force. These parameters reproduce the accuracy of \cite{Barack-Sago} to all significant figures for all data points. Our results for $a^+_{6,6L}$ are very inaccurate predictions for the corresponding PN parameters, as our fit ignores all higher--order terms.}
\begin{center}
\begin{tabular}{|c|c||c|c||c|c||c|c||}\hline\hline
$a^-_0$ & 4.57583 & $a^+_6$ & 331.525 & $b^-_0$ & 1.32120 & $b^+_0$ & 1.999991\\
\hline
$a^-_1$ & 31.8117 & $a^+_{6L}$ & -2081.57 & $b^-_1$ & 1.2391 & $b^+_1$ & -6.9969\\
\hline
$a^-_2$ & -267.250 & & & $b^-_2$ & -1.297 & $b^+_2$ & 6.29\\
\hline
$a^-_3$ & 1049.27 & & & $b^-_3$ & 1.07 & $b^+_3$ & -24.6\\
\hline\hline
\end{tabular}
\end{center}
\label{default}
\end{table}%

The simplicity of our model allows us to easily separate the CSF effects. Specifically, for quasi--circular Schwarzschild orbits we may write 
$f^{\rm SF}_{\mu}=
f^{\rm SF}_{t}
\,\delta^t_{\mu}+f^{\rm SF}_{\varphi}\,\delta^{\varphi}_{\mu}+f^{\rm SF}_{r}\,\delta^t_{r}$ where the last term on the right hand side (RHS) is purely conservative, and the first two are purely dissipative. We may therefore study the conservative effects by turning off by hand the last term on the RHS. 

\begin{figure}[h]
 \includegraphics[width=7.5cm]{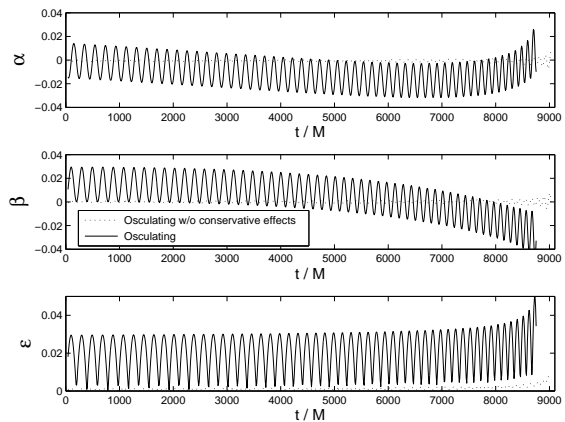}
\caption{The osculating--code variables $\alpha$ and $\beta$ as functions of the time $t$ (upper two panels), and the effective eccentricity $\epsilon$ of the orbit (lower panel) as a function of $t$ in the osculating case. We show the results from the osculating orbits method with the inclusion of the CSF (solid curves) and with turning off the CSF (dotted curves). In all cases shown here $\eta=10^{-2}$ and $p_0=10$.}
\label{alpha_beta}
\end{figure}

We integrate the SF driven orbit using two independent methods: one method is the osculating orbit approach \cite{Pound-Poisson} [specifically Eqs.~(43)--(47) therein], with special care given to the requirement that the orbit is quasi--circular. Specifically, free evolution may take the orbit away from quasi--circularity because 
integration using the osculating geodesics method cannot keep the value of the eccentricity as precisely zero. As both variables $\alpha$ and $\beta$ (see \cite{Pound-Poisson} for definitions) are dynamical, the eccentricity $\epsilon$ must evolve along the orbit too. This behavior is shown in Fig.~\ref{alpha_beta}. Interestingly, the inclusion of the conservative piece of the self force amplifies the resulting eccentricity.

\begin{figure}
 \includegraphics[width=7.5cm]{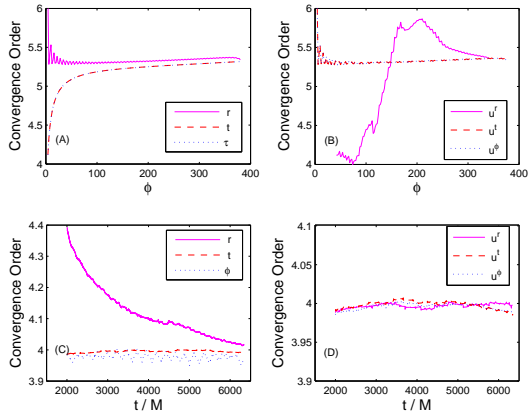}
\caption{Convergence tests for the two codes. Top panels: the 4-position (A) and 4-velocity (B) for the osculating code as functions of the azimuthal angle $\varphi$. Lower panels: the 4-position (C) and 4-velocity (D) for the direct code and a function of time $t$. In all cases shown here $\eta=10^{-2}$ and $r_0=10M$, and the CSF is included.}
\label{fig_conv}
\end{figure}

The second method is the direct integration of the orbit. The direct integration method takes the local equation of motion to be $u^{\beta}\,\nabla_{\beta}u^{\alpha}=\mu^{-1}\,f^{\alpha}_{\rm SF}$ (``Newton's second law,'' with covariant differentiation compatible with the background metric) and integrates its solution. Both codes are numerically stable and convergent. Specifically, the osculating code converges with 5$^{\rm th}$ order, and the direct code converges with $4^{\rm th}$ order (Fig.~\ref{fig_conv}).

Comparing the two independent methods for finding the SF driven waveforms is not trivial because of a difficulty in finding identical initial conditions. Specifically, the osculating method requires as initial data only the specification of the initial position vector, which in our case is taken to be a circular geodesic at some initial $p_0$. The direct method, however, requires both the position and the velocity vectors to be specified, such that the constraint equation $u_{\mu}u^{\mu}=-1$ is also satisfied. The main difficulty is that in the initial data for the osculating method the initial $u^r_0=0$, such that it corresponds to an incorrect initial radial velocity for the direct method. In practice, we find the initial data for the direct code by generating the orbital parameters at $p_0$ by running the osculating code from some $p\gg p_0$ down to $p_0$, and then take the position and velocity vectors at $p_0$ as the initial data for the direct method. The residual disagreement in the initial data can be controlled to be compatible with our numerical error tolerance.  We then use the obtained orbits to generate the waveforms using a code for the sourced Teukolsky equation with hyperboloidal slicing, a code which converges at 2$^{\rm nd}$ order \cite{code}.

As noted above, our approximation holds only for as long as $\varepsilon\gg \mu^{-1/2}$. We therefore do not integrate the equations of motion in practice all the way down to the ISCO at $p=6$, and stop the integration at a finite distance from the ISCO. In practice, we stop the integration at $p_{\rm final}=6.15\pm0.10$. Stopping at $p_{\rm final}$ is enough to estimate the dephasing at the ISCO: the dephasing $\,\Delta\Psi$ is a smooth function of the time $t$ along the particle's world line. In practice, we extrapolate $r$ as a function of $t$ to the ISCO to determine the time at which the particle arrives at the ISCO, and then we extrapolate the phase of the waveform to the same value of the time to estimate the phase of the waveform when the particle arrives at the ISCO. We may then find the difference of the total phases between two waveforms to find the dephasing $\,\Delta\Phi$. 

\section{Results}\label{results}
\subsection{The orbit}\label{orbit}

\begin{figure}[h]
 \includegraphics[width=7.5cm]{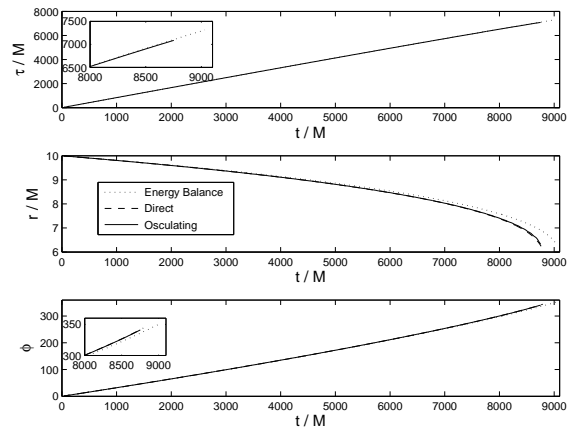}
\caption{The Orbit. The 4--position for three orbital evolution codes: energy balance (dotted), direct evolution (dashed), and osculating code (solid).}
\label{orbit1}
\end{figure}

\begin{figure}[h]
 \includegraphics[width=7.5cm]{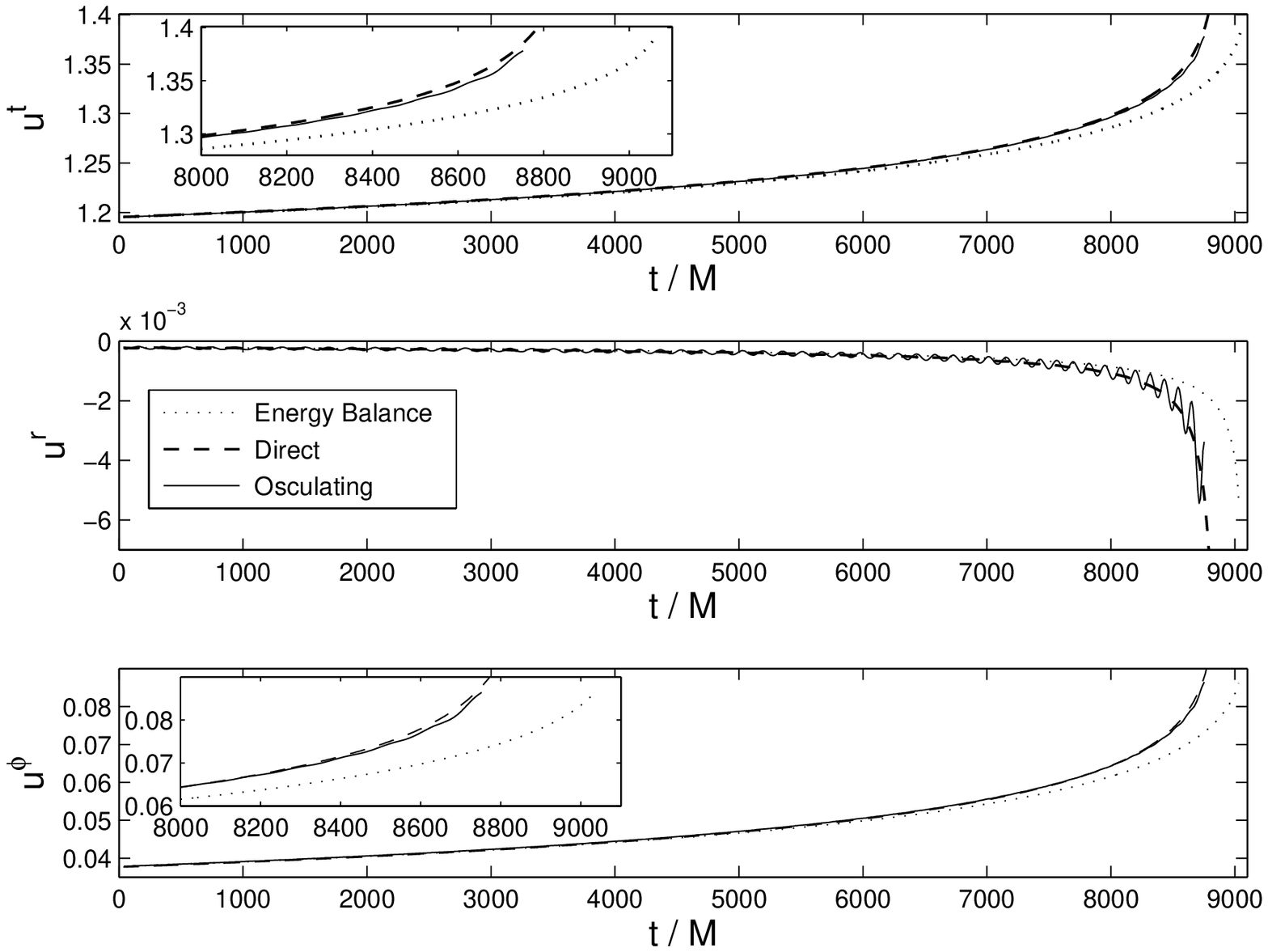}
\caption{The Orbit. The  4--velocity  for three orbital evolution codes: energy balance (dotted), direct evolution (dashed), and osculating code (solid).}
\label{orbit2}
\end{figure}

\begin{figure}[h]
 \includegraphics[width=7.5cm]{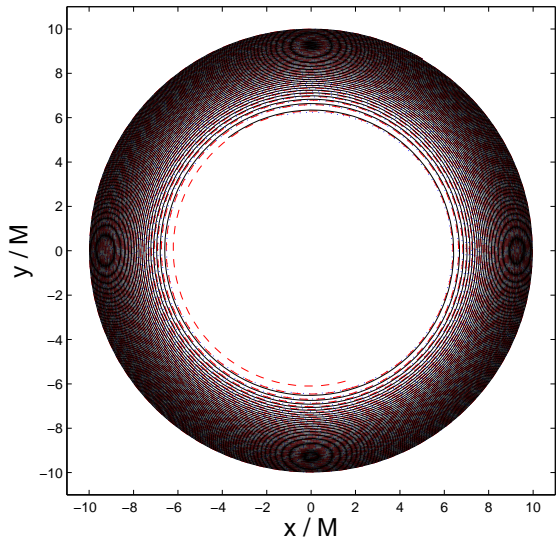}
\caption{The Orbit. The shape of the orbit for the three orbital evolution codes: energy balance (dotted), direct evolution (dashed), and osculating code (solid). }
\label{orbit3}
\end{figure}

\begin{figure}[h]
 \includegraphics[width=7.5cm]{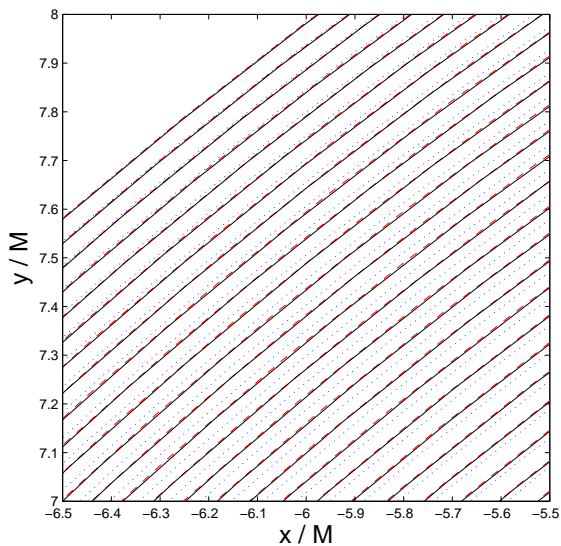}
\caption{The Orbit. The shape of the orbit for the three orbital evolution codes: energy balance (dotted), direct evolution (dashed), and osculating code (solid). A small portion of the orbit shown in Fig.~\ref{orbit3} is magnified to show detail.}
\label{orbit4}
\end{figure}

We next choose $\eta=10^{-2}$ and $p_0=10$. There is of course nothing special about this choice of the parameters,  except that we couldn't make a much higher choice for $\eta$ and  justify it  with the linearized approximation used. Below we study the dependence of the effect on the parameters $\eta,p_0$.  
The orbit is displayed in Figs.~\ref{orbit1},\ref{orbit2},\ref{orbit3} and \ref{orbit4} for the three codes. Notably, the two independent self force codes reproduce the orbit to high level of agreement, with a difference much smaller than the difference between either and the orbit generated in the energy balance approach. This difference is attributed to the effect of the conservative piece of the self force. 

\begin{figure}
 \includegraphics[width=7.5cm]{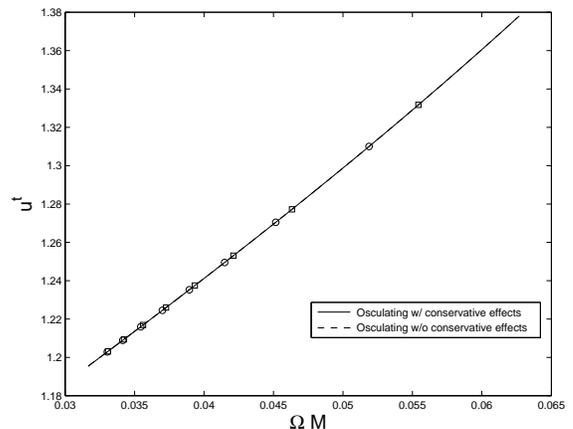}
\caption{The gauge invariant figure of $u^t$ as a function of the angular frequency $\Omega$. The curve without (dashed, $\circ$) and with the conservative effects  (solid, $\square$) are shown, together with equal-$t$ increment marks starting at $t=1,500\,M$ (at the bottom left) in increments of $1,000\,M$. The two curves are indistinguishable to the numerical accuracy of our computation (using the osculating orbits code in both cases).}
\label{gaugeinv}
\end{figure}

The orbit, of course, is a gauge dependent quantity. Indeed, the position vector changes trivially under gauge transformations, $x^{\mu}\to x^{\mu}+\xi^{\mu}$. We can, however, create gauge invariant quantities in a specific gauge choice (in our case, the Lorenz gauge), and then those quantities are guaranteed to remain unchanged in any other gauge. Two independent gauge invariant quantities are $u^t$ (``gravitational redshift", ``helical Killing vector of the perturbed spacetime") and the angular frequency $\Omega$ \cite{detweiler:2008}. In Fig.~\ref{gaugeinv} we plot $u^t$ as a function of $\Omega$ with and without the conservative piece of the self force. Notably, to the accuracy of our numerical computation the two curves overlap. That is, we find -- as expected -- that $u^t$ as a function of $\Omega$ is insensitive to the conservative piece of the self force \cite{barack:2009}. This conclusion implies that when an actual data stream is used and this gauge invariant figure is plotted, one may use a simplified radiation--reaction scheme, that does not include the conservative effects in its analysis. 

There is, however, an aspect of the gauge invariant figure (\ref{gaugeinv}) that is sensitive to the conservative effects, specifically the speed with which the data point moves along the curve. The way the conservative effects are manifested in the gauge invariant plot is not is the shape of the curve, but in the time it takes the signal to move along it. One may therefore observe the CSF effect by monitoring the motion of the data point representing the system along its curve on the $u^t-\Omega$ plane.  
We note that one additional effect of the CSF is the shift in the ISCO \cite{barack-sago:2009}, which we do not consider here.

\subsection{The waveforms}\label{waveform}

\begin{figure}
 \includegraphics[width=7.5cm]{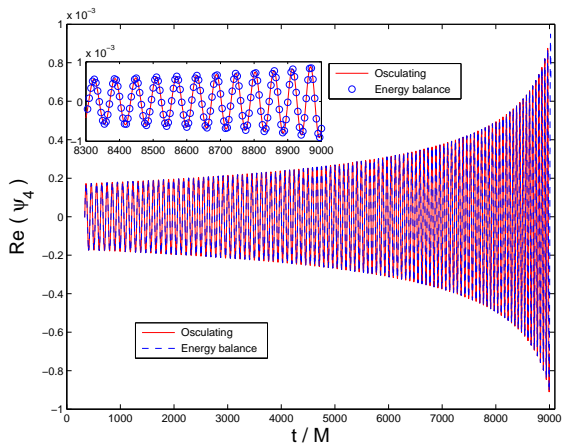}
\caption{The real part of the waveform $\psi_4$ when the CSF is turned off (solid), and using the energy balance approach (dashed, $\circ$). 
The SF waveform was calculated with the orbit evolving using the osculating method \cite{Pound-Poisson}.}
\label{fig1}
\end{figure}

We show the waveform for the case that the CSF is turned off in Fig.~\ref{fig1} for the same parameters discussed above, specifically $p_0=10$ and $\eta=10^{-2}$. We find that the waveforms obtained with the SF keeping only its dissipative pieces (and turning off its CSF) overlap with the energy balance waveform. Notice that the two waveforms in the figure are indistinguishable. The calculation method is very different in these two cases: In the SF case the orbital evolution is local; the orbit evolves because of a local force acting on the particle; in the energy balance case the fluxes to infinity and down the event horizon are calculated, and then the energy escaping over a period of the orbit is removed from the particle, and a new orbit with the new values of the constants of motion is found. The agreement of the waveforms is therefore a non-trivial test of the correctness of the calculation. The two waveforms overlap nearly exactly, with total cumulative dephasing at the order of $10^{-3}$ radians. 

\begin{figure}[h]
 \includegraphics[width=7.5cm]{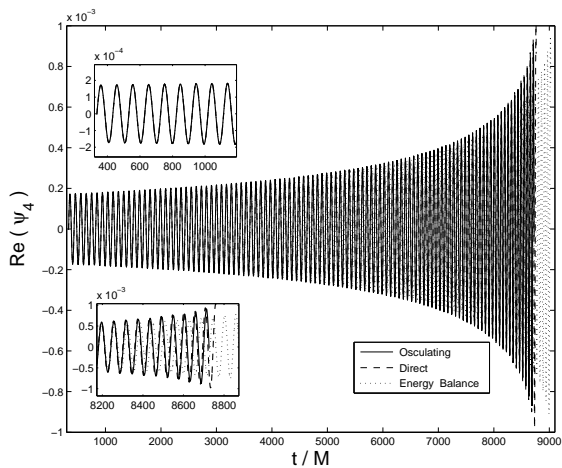}
\caption{The real part of the waveform $\psi_4$ when the CSF is included (solid for the osculating orbit method, and dashed for the direct integration method), and using the energy balance approach (dotted). The three waveforms are in phase at the beginning of the waveforms, but the former two dephase with time from the latter (and to a much smaller extent from each other).}
\label{fig2}
\end{figure}

We next reintroduce the CSF. We integrate the SF driven orbit using two independent methods: the osculating orbit approach and direct integration of the orbit. Figure \ref{fig2} shows the waveform for the energy balance approach (same as in Fig.~\ref{fig1}) and for the two independent methods of calculating the SF driven orbit (including the CSF). The latter two waveforms are in agreement with each other with small dephasing (see below) between them, and a much larger dephasing of either with the energy balance waveform.

\begin{figure}
 \includegraphics[width=8.0cm]{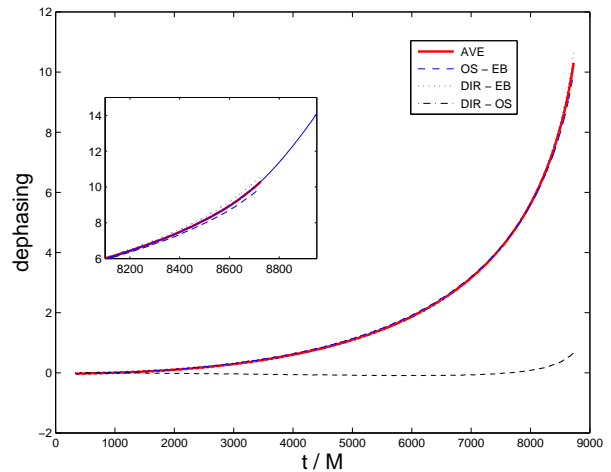}
\caption{The dephasing of the waveforms for the same data shown in Fig.~\ref{fig2}. We show the dephasing from the energy balance waveform of the waveforms obtained from both methods to calculate the orbital evolution when the conservative effects are included, and the dephasing of the two latter methods from each other.  Dashed curve: 
dephasing of the waveforms from the osculating orbit method from the energy balance method; Dotted curve: dephasing of the waveforms from the direct integration method from the energy balance method; Thick solid curve: the average dephasing of the waveform. Dash--dotted curve: The difference between the dotted and dashed curves. 
Insert: same as in the main figure, with the extrapolated dephasing down to the ISCO shown in a thin solid curve.}
\label{fig3}
\end{figure}

The waveform dephasing of Fig.~\ref{fig2} is shown explicitly in Fig.~\ref{fig3}. We find that the total cumulative dephasing of the waveforms at the endpoint of the evolution at $p_{\rm final}$ is $\,\Delta\phi=10.3\pm0.4$ radians. The error estimate comes from the numerical errors in each calculation method and from residual incompatibility of initial data in the two SF driven cases, which we estimate by comparing the waveforms obtained from the osculating orbit approach and the direct integration approach. 

The dependence of $\,\Delta\Phi$ on the position $p$ is a simple monotonic unction of the time (Fig.~\ref{fig3}) which we can extrapolate from $p_{\rm final}$ down to the ISCO at $p=6$. We find that at the ISCO the dephasing is $\,\Delta\Phi=14\pm 1$ radians. 
Dephasing of $14\pm 1$ radians corresponds to about $2.2$ cycles over the entire motion of the particle over 107.8  cycles. We next consider the following simulation of a detection event. Say the actual data stream is modeled by the waveforms obtained with the full SF expression, i.e., including the CSF, and that the theoretical template is obtained by turning off the CSF, or equivalently by using the energy balance waveform. By how much is the overlap integral of the waveforms reduced because of our ignorance of the CSF? 

\begin{figure}
 \includegraphics[width=7.0cm]{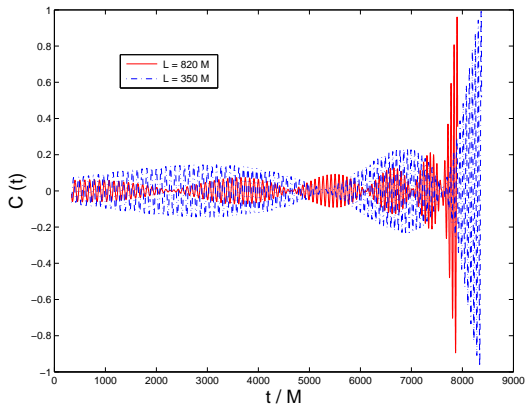}
\caption{The local overlap integral for a window of length $L=350M$ (dash-dotted) and of length $L=820M$ (solid) taken from the energy balance waveform and a piece of equal length of the full SF (including conservative effects) waveform obtained from the osculating orbit method, as a function of the starting point of the latter. The window is taken here from the end of the waveform (the late chirp part). The global maximum of the overlap integral is 0.9900 (for $L=350M$) and 0.9594 (for $L=820M$).}
\label{fig4}
\end{figure}

\begin{figure}
 \includegraphics[width=7.0cm]{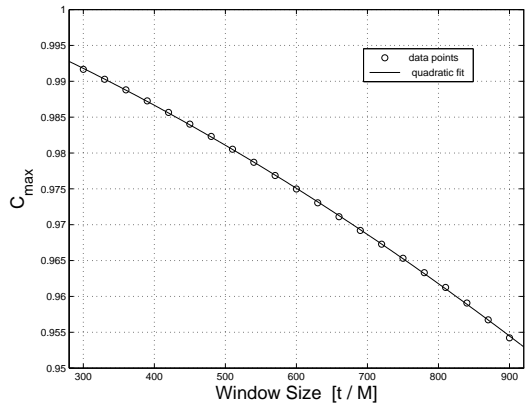}
\caption{The maximal overlap integral for windows of varying length $L$ taken from the energy balance waveform and a piece of equal length of the full SF (including conservative effects) obtained from the osculating orbit method, as a function of the starting point of the latter. The circles are data points, and the curve is the quadratic fit 
$C_{\rm max}=1.0010-1.7988\times 10^{-5}\,L-3.9154\times 10^{-8}\,L^2$, with fit parameter $R^2=0.9983$.}
\label{fig5}
\end{figure}

Figure \ref{fig4} shows the overlap integral as a function of the time, when we take a window of length $L$ of the energy balance waveforms (specifically from their late chirp part), and shift it along the waveform of the full SF. At each point we calculate the overlap integral of the window with a local piece of the second waveform, and plot the local overlap integral as a function of the start time of the window. Because the window was taken from the late part of the waveform, we find that the overlap integral is very small at first, and becomes large only at the late part of the other waveform. We then take the maximum of the overlap integral to be the one corresponding to the chosen window $L$.  More precisely, we calculate
$C_{\rm max}={\rm max}_{\tau}\,\frac{<\psi^{\rm SF}(t) | \psi^{\rm EB}(t-\tau)>}
{\sqrt{<\psi^{\rm SF}(t) | \psi^{\rm SF}(t)><\psi^{\rm EB}(t) | \psi^{\rm EB}(t)>}}\, .$
Larger values of $L$ would reduce the overlap integral even further. In Fig.~\ref{fig5} we show $C_{\rm max}$ as a function of the window size $L$. 
As $L$ increases, at some value $C_{\rm max}$ would drop below a pre-determined value that marks our tolerance for detection or parameter estimation. Many times this threshold is taken to be $C=0.96$, because then detection rate would drop by $10\%$. Here, this threshold is obtained when $L_{\rm threshold}=816.6\,M$, which corresponds to just over 14 wavelengths of the emitted GW. If $L\gtrsim L_{\rm threshold}$, the exclusion of the CSF effects would cause a significant drop in the $C_{\rm max}$ that would reduce the detection rate by $10\%$ or more.  In such a case ignoring the CSF would have an important effect on detection or parameter estimation of the GW. However, short waveforms (i.e., $L < L_{\rm threshold}$) do not require the CSF effects to be included if reduction of the detection rate by less than this tolerance is acceptable.  
The overlap integral increases rapidly as the template window is taken from earlier times. 
This result suggests that when other parts than the very end  of the waveform is of interest, the full SF is even less significant than we have found. 

\subsection{Varying the values of parameters}\label{varying}

Our model of quasi--circular Schwarzschild orbits depends on two variables: the mass ratio $\eta$ and the start point $p_0$. Here we vary each parameter independently and find the dependence of the dephasing $\,\Delta\Phi$ [between the osculating orbits case (that includes the CSF) and the energy balance case (that neglects the CSF and considers only dissipative effects)] on either parameter. 

\subsubsection{Varying the mass ratio $\eta$}

First we study the variation of the dephasing with changing the parameter $\eta$, the mass ratio. The greatest problem with varying $\eta$ is its effect on the computation time. On the one hand we cannot justifiably increase $\eta$ beyond $10^{-2}$, because then the linearization approximation breaks down. On the other hand, lowering $\eta$ to very small values, while satisfying the linearization requirement more confidently, results in longer physical evolution times and correspondingly also longer computation times. 

In practice we reduce the value of $\eta$ by a full order of magnitude, and sample values in the range $[10^{-3}-10^{-2}]$, and fix $p_0=8$. (We decrease the value of $p_0$ from its previous value of 10 to save on computation time for the lower values of $\eta$.) 

\begin{figure}
 \includegraphics[width=7.0cm]{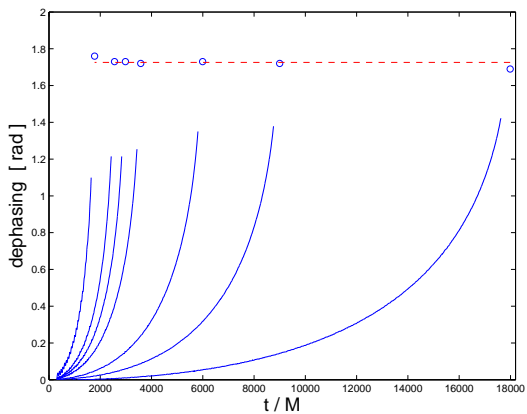}
\caption{The dephasing $\,\Delta\Phi$ as function of time for a family of mass ratios $\eta$ for the values (from right to left): $\eta=$0.001, 0.002, 0.003, 0.005, 0.006, 0.007, and 0.010 (solid curves). The circles display the extrapolated values when the particle arrives at the ISCO, and the dashed line shows their average at $1.726\pm 0.021$. In all cases the motion starts at $p_0=8$.}
\label{fig6}
\end{figure}

\begin{figure}
 \includegraphics[width=7.0cm]{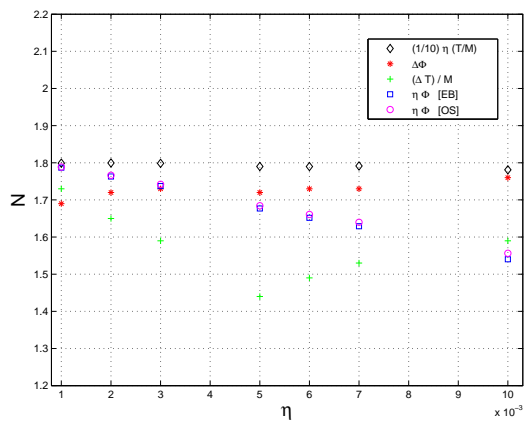}
\caption{Five dimensionless quantities (denotes collectively by $N$) as functions of the mass ratio $\eta$: $\mu T/M^2$ (divided by 10 to keep the scale similar with the other four quantities) ($\diamond$), the dephasing $\,\Delta\Phi$ ($*$), the difference in arrival time $\,\Delta T/M$, and $\eta\Phi$ for the energy balance case ($\square$) and the osculating orbits case ($\circ$). Here, $\,\Delta\Phi$ and $\,\Delta T/M$ are between the osculating orbits and the energy balance cases, and $T$ is the total time of motion from $p_0$ down to the ISCO.  In all cases the motion starts at $p_0=8$ and the values shown are extrapolations to the ISCO.}
\label{fig7}
\end{figure}

Figure \ref{fig6} shows a family of curves displaying the dephasing between the cases of the osculating orbits method  and the energy balance method. 
Each curve in Fig.~\ref{fig6} ends at the point we stop the integration when the orbits gets too close to the ISCO for the adiabatic approximation to still hold. These curves are very smooth and simple functions, which we extrapolate to the times at which the particle arrives at the ISCO. The extrapolated value of the dephasing at the ISCO is also shown in Fig.~\ref{fig6}.  We find only little variation in the dephasing as $\eta$ changes over a full order of magnitude, consistent with the expectation that the dephasing has only a weak dependence on $\eta$. Indeed, one could expect from scaling arguments that the dephasing per orbit scales with $\eta$ while the number of orbits scales with $\eta^{-1}$ (both scalings are indeed found in our simulations --- see Fig.~\ref{fig7}), so that the total dephasing is at the leading order at $O(\eta^0)$. Here we show that not only is the dephasing at $O(1)$, but in the range tested and with our numerical resolution is indistinguishable from a constant value, or at the most is a very weak function of $\eta$. 

In Fig.~\ref{fig7} we show five dimensionless quantities constructed from the evolution time $T$ from $p_0$ down to the ISCO, the total phase of the waveform in the energy balance and osculating methods case (including the CSF term), and the differences in arrival time and the dephasing between the latter two. We find that $T\sim O(\eta^{-1})$, that $\,\Delta\Phi\sim O(\eta^{0})$, $\,\Delta T/M\sim O(\eta^{0})$ and that in  both the energy balance and osculating orbits cases the total phase $\Phi$ is a linear function of $\eta$ with a very small slope, that is at the magnitude of our computational error. We therefore cannot rule out that we see in addition to the leading $O(\eta^{-1})$ term also a higher-order $O(\eta^{0})$ term. Notice that the last four quantities are comparable to each other, and that the variation in all five (at the most $10\%$) is very small compared with the full order of magnitude variation in $\eta$.

\subsubsection{Varying the initial position $p_0$}

\begin{figure}
 \includegraphics[width=7.0cm]{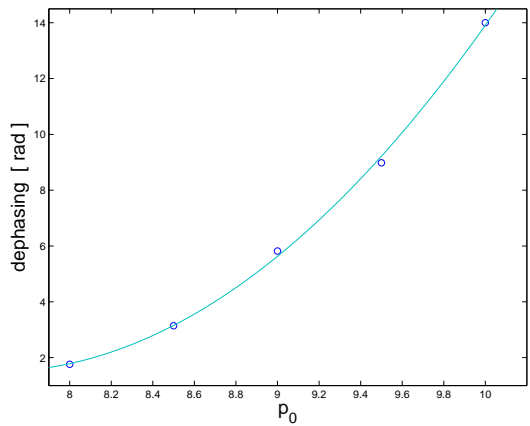}
\caption{The dephasing $\,\Delta\Phi$ as function of the initial position $p_0$. The circles display the extrapolated values when the particle arrives at the ISCO, and the curve shows a quadratic fit. In all cases the mass ratio $\eta=10^{-2}$.}
\label{fig8}
\end{figure}

Next we fix the mass ratio $\eta$ and vary the starting point of the motion $p_0$. In practice we choose the value of  $\eta=10^{-2}$. Figure \ref{fig8} shows the dephasing between the osculating orbits case (that includes the CSF) and the energy balance case (that neglects the CSF) for the range $p_0\in[8,10]$. Naturally, the dephasing grows with $p_0$. The data presented are consistent a quadratic dependence of $\,\Delta\Phi$ on $p_0$.  

After the completion of this work we became aware of Warburton {\it et al} \cite{warburton}. The approach of \cite{warburton} is similar to our osculating method, except that \cite{warburton} estimates the dephasing by the difference in the azimuthal angle $\varphi$ of the orbit, whereas we actually compute the waveforms and find their dephasing. The generalization of our quasi--circular orbit to bound orbits of varying eccentricity --- as is done in \cite{warburton} --- is straightforward, as the osculating orbits equations of motion already include the eccentricity parameter. Lastly, we compute the waveforms using two independent computational methods, specifically the osculating method and the direct method. The direct method does not appear to us to be convenient for  generalization to generic Kerr orbits, and the osculating method has a clear advantage over it for such orbits.

\section*{Acknowledgments}

The authors wish to thank Gaurav Khanna for discussions and for use of his numerical code. 
This work has been supported by a NASA EPSCoR RID grant and by NSF grants PHY--0757344, PHY--1249302 and DUE--0941327. LMB is grateful to Alessandro Spallicci for hospitality. 

\begin{appendix}

\section{The 5.5 Post--Newtonian term}\label{PN}

The ${\rm PN}_{5.5}$ term used in Eq.~(\ref{pn}) is given by (\cite{tanaka:1996}):

\begin{eqnarray*}
{\rm PN}_{5.5}&=& 1 - \frac{1247}{336} \Bigg(\frac{M}{r}\Bigg) + 4 \pi\Bigg(\frac{M}{r}\Bigg)^{\frac{3}{2}} - 
 \frac{44711}{9072}\Bigg(\frac{M}{r}\Bigg)^2 \\
 &-& \frac{8191}{672}\ \pi \Bigg(\frac{M}{r}\Bigg)^{\frac{5}{2}}+  
\Bigg( \frac{6643739519}{69854400} 
 - \frac{1712}{105}\,  \gamma  
+  \frac{16}{3}\  \pi^2\\ 
&-& \frac{3424}{105}\  \ln 2 
 -\frac{856}{105} \,\ln\frac{M}{r}\Bigg)\Bigg(\frac{M}{r}\Bigg)^3 
   - \frac{16285}{504} \pi \Bigg(\frac{M}{r}\Bigg)^{7/2} \\
 &+& \Bigg(-\frac{323105549467}{3178375200}+ 
\frac{232597}{4410}\, \gamma - \frac{1369}{126}\  \pi^2 \\
&+& \frac{39931}{294}\  \ln 2 - \frac{47385}{1568}\  \ln 3  +  
  \frac{232597}{8820}\ \ln \frac{M}{r}\Bigg) \Bigg(\frac{M}{r}\Bigg)^4\\
  &+& \pi\, \Bigg(\frac{265978667519}{745113600} - \frac{6848}{105}\, \gamma
-  \frac{13696}{105}\, \ln 2 \\
&-&  \frac{3424}{105}\ \ln \frac{M}{r}\Bigg) \Bigg(\frac{M}{r}\Bigg)^{9/2}
+\ \Bigg(-\frac{2500861660823683}{2831932303200} \\
&+& \frac{916628467}{7858620}\ \gamma
- \frac{424223}{6804}\,\pi^2 
- \frac{83217611}{1122660}\ \ln 2\\
&+& \frac{47385}{196}\  \ln 3
+  \frac{916628467}{15717240}\ \ln\frac{M}{r}\Bigg) \Bigg(\frac{M}{r}\Bigg)^{5} \\
&+&\pi\, \Bigg(\frac{8399309s750401}{101708006400}
+ \frac{177293}{1176}\ \gamma
+  \frac{8521283}{17640} \ \ \ln 2\\
&-& \frac{142155}{784} \ \ \ln 3 
+  \frac{177293 }{2352}\  \ln\frac{M}{r}\Bigg)\Bigg(\frac{M}{r}\Bigg)^{11/2}
\end{eqnarray*}

\end{appendix}

\end{document}